\documentclass[twocolumn,showpacs,prl,amsmath,amssymb,superscriptaddress]{revtex4}
\usepackage{graphicx}% Include figure files
\usepackage{dcolumn}% Align table columns on decimal point
\usepackage{bm}% bold math
\usepackage{amssymb}

\begin{document}

\title{Superconductors as spin sources for spintronics}

\author{F. Giazotto}
\email{giazotto@sns.it}
\affiliation{NEST CNR-INFM and Scuola Normale Superiore, I-56126 Pisa, Italy}
\author{F. Taddei}
\affiliation{NEST CNR-INFM and Scuola Normale Superiore, I-56126 Pisa, Italy}
%\date{\today}

\begin{abstract}
Spin-polarized transport is investigated in normal metal-superconductor (NS) junctions as a function of interface transmissivity as well as temperature when the density of states of a superconductor is Zeeman-split in response to an exchange field ($h_{\text{exc}}$).
Similarly to the "absolute spin-valve effect" predicted by D. Huertas-Hernando \emph{et al}. [Phys. Rev. Lett. \textbf{88}, 047003 (2002)] in superconducting proximity structures, we show that NS junctions can be used to generate highly spin-polarized currents, in alternative to half-metallic ferromagnets.
In particular, the spin-polarized current obtained is largely tunable in magnitude and sign by acting on bias voltage and $h_{\text{exc}}$. While for tunnel contacts the current polarization can be as high as $100\%$, for transparent junctions it is dominated by the minority spin species.
The effect can be enhanced by electron "cooling" provided by the superconducting gap.
\end{abstract}


\maketitle
 
\emph{Spintronics} is a research field where two fundamental branches of physics, i.e., magnetism and electronics, are combined \cite{fabian,aws}, and it is usually based on the opportunity of ferromagnetic materials to provide spin-polarized currents \cite{johnson,jedema,urech,huang,jonker,giazotto03}. The effectiveness of spintronics depends on the extent to which a current is spin polarized, which turns out to depend on the degree of polarization of the ferromagnet (F). The performance of any spintronic device, in fact, improves as the polarization approaches 100\%, a condition achievable through the exploitation of half-metallic ferromagnets~\cite{fabian}.
The availability of highly spin-polarized sources is thus of crucial importance from both fundamental and technological side~\cite{datta}.

In Ref. \cite{nazarov} D. Huertas-Hernando \emph{et al}. showed that an "absolute spin-valve effect" and 100\% current polarization can be obtained in superconducting proximity tunnel structures composed of two coupled trilayers. These consist of a normal layer tunnel coupled to a ferromagnetic layer, on one side, and a superconducting layer, on the other, so that in the normal region superconducting and magnetic correlations are induced producing a spin-split BCS-like density of states. The polarized current is found through the two tunnel-coupled N layers.
In this Brief Report, we theoretically address spin polarized transport in normal metal-superconductor (NS) junctions as  a function of interface transmissivity and temperature showing that superconductors can be used to produce highly-polarized spin currents provided the Zeeman interaction dominates their response.
We show, on the one hand, that 100\% spin polarization is achievable in the \emph{tunnel} limit, and, on the other, that the polarization is largely tunable in sign and magnitude by acting on the bias voltage and on the Zeeman energy.
Remarkably, for perfectly \emph{transparent} NS interfaces, the current is dominated by the \emph{minority} spin species. Furthermore, the effect of the temperature is to smear and suppress the polarization.
The efficacy of this method, enhanced by the \emph{cooling} effect occurring in NS interfaces, makes Zeeman-split superconductors prototype candidates, alternative to half-metallic ferromagnets, for low-temperature spintronics \cite{prl05,apl06}.   
\begin{figure}[t!]
\includegraphics[width=\columnwidth,clip]{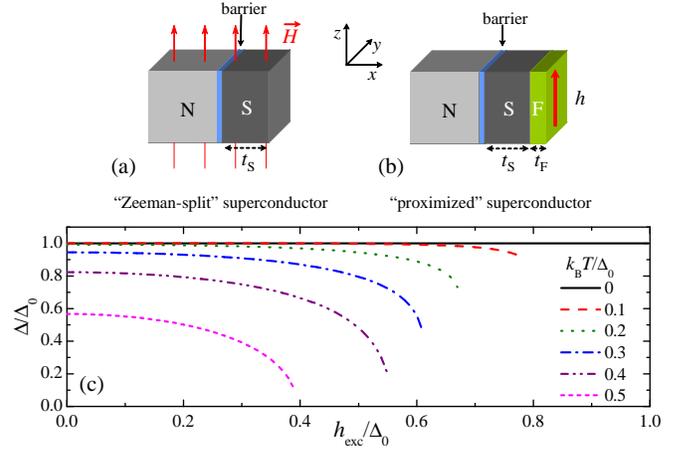}
\caption{(color online)  The system under investigation consists either of (a) a NS junction in a static magnetic field $H$ applied parallel to the interface (i.e., the standard "Zeeman-split" superconductor configuration), or (b) a NSF junction where an exchange field is induced in the superconductor through \emph{proximity} effect induced by a nearby ferromagnetic layer in good electric contact with S. The F exchange field ($h$) is confined to the $y-z$ plane.
(c) Order parameter $\Delta$ vs exchange field $h_{\text{exc}}$ calculated at different temperatures.}
\label{fig1}
\end{figure}

The system under investigation consists equivalently of a three-dimensional ballistic NS  junction in a static in-plane magnetic field $H$ [see Fig. \ref{fig1}(a)], or a NSF trilayer [see Fig. \ref{fig1}(b)] where the exchange field ($h$) is provided by a ferromagnetic film in good electric contact with the superconductor \cite{proximity}.
More precisely, the normal metal-superconductor junction consists of a contact whose transverse dimensions are much smaller than the elastic  mean free path in N and S, i.e., a Sharving ballistic contact.
The S interface is located at $x=0$, and electron transport occurs along the $x$ direction.
The first setup requires a very thin film, $t_{\text{S}}\ll \lambda$, where $t_{\text{S}}$ is the superconductor thickness and $\lambda$ is the magnetic penetration length, so that a Zeeman energy $h_{\text{exc}}=\frac{1}{2}g\mu_{\text{B}}H$, where $\mu_{\text{B}}$ is the Bohr magneton and $g$ is the gyromagnetic factor \cite{meservey}, is induced while orbital effects are negligible.
In the second setup, if $t_{\text{S}}$ is smaller than the superconducting coherence length and the F thickness ($t_{\text{F}}$) is smaller than the length of the condensate penetration into the ferromagnet, the influence of the F layer on the superconductor becomes \emph{nonlocal}, and the ferromagnet induces in S a homogeneous \emph{effective} exchange field ($h_{\text{exc}}$) through proximity effect, thus modifying the superconducting gap ($\Delta$) \cite{Bergeret}. As stated in Ref. \cite{Bergeret}, $h_{\text{exc}}$ is much smaller than $h$ and of the same order of magnitude as the modified gap. 

In order to study the electron transport in the structure we use the Bogolubov-de Gennes equation \cite{de Gennes} which, in the absence of spin-flip scattering, reads \cite{dejong}
\begin{equation}
 \begin{pmatrix}
  H_{0}-\sigma h_{\text{exc}} &  \Delta(T,h_{\text{exc}})\\
  \Delta(T,h_{\text{exc}})        & -H_{0}-\sigma h_{\text{exc}}
 \end{pmatrix}
 \begin{pmatrix}
  u_{\sigma}\\
  v_{-\sigma} 
 \end{pmatrix}
 = \varepsilon
 \begin{pmatrix}
  u_{\sigma}\\
  v_{-\sigma}
 \end{pmatrix},
\label{eq:ehSm}
\end{equation}
where $H_0$ is the single-particle Hamiltonian,
\begin{eqnarray}
u_{\sigma}^2=\frac{1}{2}\left[1+\frac{\sqrt{(\varepsilon+\sigma h_{\text{exc}})^2-\Delta(T,h_{\text{exc}})^2}}{\varepsilon+\sigma h_{\text{exc}}}\right],\\
v_{-\sigma}^2=\frac{1}{2}\left[1-\frac{\sqrt{(\varepsilon+\sigma h_{\text{exc}})^2-\Delta(T,h_{\text{exc}})^2}}{\varepsilon+\sigma h_{\text{exc}}}\right]
\end{eqnarray}
are the BCS coherence factors,
$\sigma =\pm 1$ is the spin, and $T$ is the temperature. The excitation energy $\varepsilon$ is measured from the condensate chemical potential $\mu$. We consider a $\delta$-like elastic scattering potential located at the NS interface  $\mathcal{V}(x)=\hbar\sqrt{2\mu/m}Z\delta(x)$ ($m$ is the electron mass), which allows to interpolate from a metallic contact ($Z=0$) to a tunnel barrier ($Z\rightarrow \infty$) \cite{BTK}. For the order parameter we use a "rigid boundary condition" [$\Delta(T,h_{\text{exc}})=\Delta(T,h_{\text{exc}})\theta(x)$, where $\theta(x)$ is the step function] which holds for a Sharving point contact \cite{likharev}.
The gap dependence on $T$ and $h_{\text{exc}}$, shown in Fig. 1(c), is determined self-consistently \cite{de Gennes,zheng} from the gap equation
\begin{eqnarray}
\text{ln}\left[\frac{\Delta_0}{\Delta(T,h_{\text{exc}})}\right]=\int_{0}^{\hbar \omega_{\text{D}}}\frac{d\epsilon}{\sqrt{\epsilon^2+\Delta^2(T,h_{\text{exc}})}}\\
\times[f_{+}(\epsilon,T,h_{\text{exc}},\Delta)+f_{-}(\epsilon,T,h_{\text{exc}},\Delta)]\nonumber,
\end{eqnarray}
where
\begin{equation}
f_{\pm}=\frac{1}{\text{exp}\left[\frac{1}{k_{\text{B}}T}\left(\sqrt{\epsilon^2+\Delta^2(T,h_{\text{exc}})}\mp h_{\text{exc}}\right)\right]+1},
\end{equation}
$\Delta_0=\Delta(0,0)$ is the zero-temperature order parameter in the absence of exchange field, and $\omega_{\text{D}}$ is the Debye frequency.
At $T=0$, $\Delta$ is independent of $h_{\text{exc}}$, suddenly dropping to zero as the exchange field equals $\Delta_0$. At finite temperatures, $\Delta$ monotonically decreases  by increasing $h_{\text{exc}}$, dropping to zero at a threshold exchange field whose value decreases by increasing $T$. Such sudden drops in the curves indicate a first-order phase transition from the superconducting to the normal state \cite{zheng}.
\begin{figure}[t!]
\includegraphics[width=\columnwidth,clip]{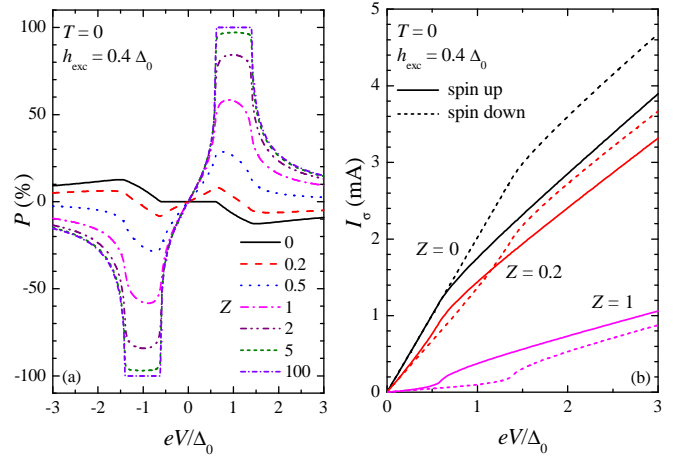}
\caption{(color online)  (a) Current polarization $P$ vs bias voltage $V$ for several $Z$ values at $T=0$ and $h_{\text{exc}}=0.4\Delta_0$. (b) Spin-dependent electric current $I_{\sigma}$ vs $V$ calculated for some $Z$ values at $T=0$ and $h_{\text{exc}}=0.4\Delta_0$.
}
\label{tS}
\end{figure}

Within the Landauer-Buttiker approach \cite{Landauer} the electric current through the junction is given by $I(V)=\Sigma_{\sigma}I_{\sigma}(V)$, where~\cite{bagwell}
\begin{eqnarray}
I_{\sigma}(V)=\frac{em\mathcal{A}\mu}{4\pi^2\hbar^3}\int_0^{\pi/2}{d\phi\, \sin{2\phi}}\int_{-\infty}^{\infty}d\varepsilon\Big(1+\frac{\varepsilon}{\mu}\Big)\nonumber\\
\times[1+R^a_{-\sigma}(\varepsilon,\phi)-R^0_{\sigma}(\varepsilon,\phi)][f_0(\varepsilon-eV)-f_0(\varepsilon)],
\end{eqnarray}
$\mathcal{A}$ is the junction area, $\phi$ is the electron injection angle in the $x-y$ plane, $f_0(\varepsilon)=[1+\text{exp}[\varepsilon/k_{\text{B}}T]]^{-1}$ is the  Fermi distribution function, $R^0_{\sigma}=|r^0_{\sigma}|^2$ ($R^a_{-\sigma}=|r^a_{-\sigma}|^2$) is the normal (Andreev) reflection probability for spin-$\sigma$ quasiparticles, and
\begin{eqnarray}
r^0_{\sigma}=\frac{(Z^2+iZ\text{cos}\phi)(v^2_{-\sigma}-u^2_{\sigma})}{u^2_{\sigma}\text{cos}^2\phi+Z^2(u^2_{\sigma}-v^2_{-\sigma})},\\
r^a_{-\sigma}=\frac{u_{\sigma}v_{-\sigma}\text{cos}^2\phi}{u^2_{\sigma}\text{cos}^2\phi+Z^2(u^2_{\sigma}-v^2_{-\sigma})}.
\end{eqnarray}
The amplitudes $r^0_{\sigma}$ and $r^a_{-\sigma}$ are obtained  through standard mode-matching along the lines of Ref. \cite{BTK}.

The spin injection properties of the superconductor can be quantified by the current polarization, defined as \cite{parameters}
\begin{equation}
P(V)=\frac{I_{+}(V)-I_{-}(V)}{I_{+}(V)+I_{-}(V)}. 
\end{equation}
Figure 2(a) displays the current polarization $P$ vs bias voltage $V$ at $T=0$ and $h_{\text{exc}}=0.4\Delta_0$ calculated for several values of $Z$.
$P(V)$ is strongly dependent on $Z$ and it is an antisymmetric function of bias voltage.
Let us consider positive voltages.
For a metallic interface ($Z=0$) $P$ is zero up to $eV=\Delta_0-h_{\text{exc}}$, thereafter becoming \emph{negative} and presenting a minimum around $eV=\Delta_0+h_{\text{exc}}$ where $P\sim -12.6\%$.
%At larger bias voltage $P$ tends asymptotically to zero.
For opaque interfaces $P$ is fully \emph{positive}: at large $Z$ values $P$ turns out to be maximized in the interval $|eV-\Delta_0|\leq h_{\text{exc}}$, reaching values as high as $100\%$ for $Z=100$ (i.e., in the tunnel limit).
For small intermediate values of $Z$ the current polarization can be both negative and positive depending on $V$.

\begin{figure}[t!]
\includegraphics[width=\columnwidth,clip]{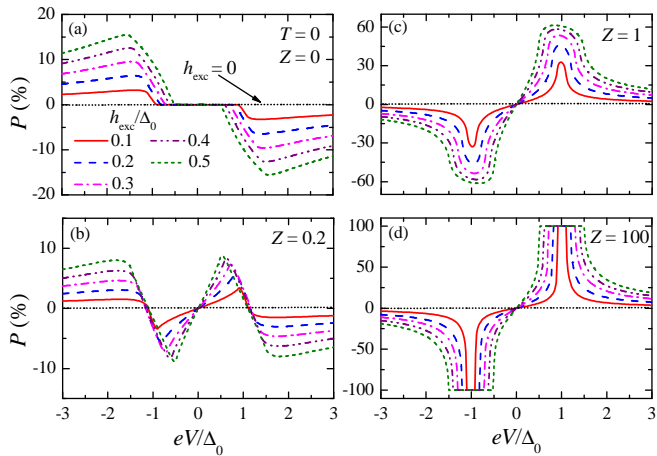}
\caption{(color online) Current polarization $P$ vs bias voltage $V$ calculated for several $h_{\text{exc}}$ at $T=0$: (a) $Z=0$, (b) $Z=0.2$, (c) $Z=1$, and (d) $Z=100$.
}
\label{tF}
\end{figure}
The behavior of $P$ for different interface transmissivities can be understood by inspecting Fig. 2(b) which shows the spin-dependent current $I_{\sigma}$ vs $V$ calculated for some relevant $Z$ values.
For $Z=0$ the current, for both spin species, takes its maximum value, since $R^a_{\pm}=1$ for all the voltages for which quasiparticle propagation in S is prohibited.
To be more precise, due to the energy shift caused by $h_{\text{exc}}$ in the spin-dependent density of states, $R^a_{-\sigma}(\epsilon,\phi)=1$ up to $\epsilon=\Delta_0-\sigma h_{\text{exc}}$ for spin $\sigma$ electrons.
Beyond this threshold quasiparticle propagation sets in, causing the differential conductance to decrease.
Since the threshold for spin up electrons ($V_+$) occurs for a smaller voltage with respect to spin down electrons ($V_-$), we have that $I_{+}<I_{-}$ so that $P$ has \emph{opposite} sign with respect to $h_{\text{exc}}$ for $eV>\Delta_0-h_{\text{exc}}$.
For opaque interfaces (i.e., $Z=1$) Andreev reflection is very much suppressed, so is the current up to voltages for which quasiparticle propagation in S takes over, and increasing thereafter.
Again, since $V_{+}<V_{-}$, the situation is now reversed and $I_{+}>I_{-}$, as expected from the tunnel-like characteristic of NS junctions with different energy gaps \cite{tinkham}.
For intermediate values of $Z$ (e.g., $Z=0.2$) the spin-dependent currents can intersect, leading to positive or negative values of $P$ depending on $V$. This non-trivial behavior originates from the crossover between Andreev-dominated transport to quasiparticle tunneling, and it is a unique property of superconductors.
  
Figure 3 displays $P$ vs $V$ calculated for several $h_{\text{exc}}$ at $T=0$ for some $Z$ values. 
For low and moderate $Z$ values ($Z\leq 1$) [see panels (a)-(c)], an increase of $h_{\text{exc}}$ leads to an enhancement of the maximum absolute value of the achievable current polarization and widens the voltage intervals of large $P$. Note, for example, that already for $Z=1$ $P$ values as high as $60\%$ can obtained for $eV\simeq \Delta_0$.
For opaque NS contacts ($Z=100$) [see panel (d)] the net effect of an increase in $h_{\text{exc}}$ is to widen the voltage regions of $100\%$ spin polarized current. From this follows that larger $h_{\text{exc}}$ as well as high $Z$ values (i.e., more tunnel-like interfaces) are preferable in order to maximize the current polarization in NS junctions.  

\begin{figure}[t!]
\includegraphics[width=\columnwidth,clip]{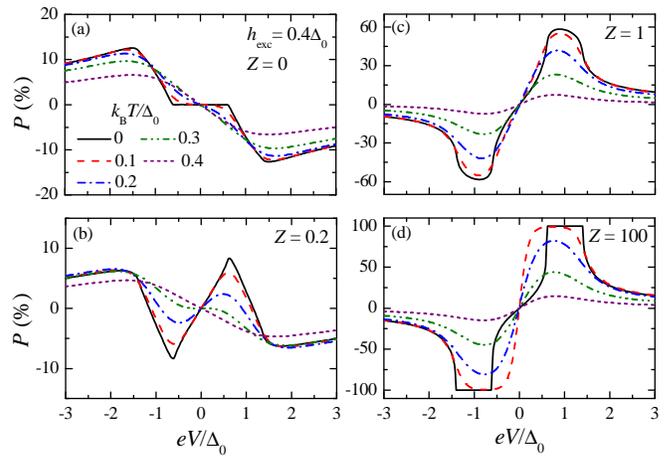}
\caption{ (color online) Current polarization $P$ vs bias voltage $V$ calculated at different temperatures $T$ 
for $h_{\text{exc}}=0.4\Delta_0$: (a) $Z=0$, (b) $Z=0.2$, (c) $Z=1$, and (d) $Z=100$.}
\label{exch}
\end{figure}
The impact of temperature $T$ on $P$ is shown in Fig. 4 for some $Z$ values and $h_{\text{exc}}=0.4\Delta_0$. For all values of $Z$, an increase of $T$ yields a smearing of the sharp features present at $T=0$, and a suppression of the maximum current polarization values.
In particular, for opaque interfaces [see panels (c) and (d) of Fig. 4], while $P$ turns out to be only marginally affected up to $k_{\text{B}}T\simeq0.1\Delta_0$, at $k_{\text{B}}T=0.3\Delta_0$ it is reduced by about $60\%$ with respect to $T=0$ . Furthermore, the maxima of $|P|$ tends to move towards zero bias at higher $T$, owing to the temperature-induced suppression of $\Delta$. These results show that $k_{\text{B}}T\ll \Delta_0$ is the condition required to obtain highly spin-polarized currents.

We shall further comment the experimental feasibility of S electrodes as pure spin injectors. In particular, the setup of Fig. 1(a) can be realized through  aluminum (Al) thin films ($\lesssim 10$ nm) tunnel-coupled to N electrodes, placed in parallel magnetic fields of the order $\sim 10^4$ Oe \cite{meservey,adams,butko}. Such a setup may suffer, however, some limitations in the case of semiconductors,  due to the additional Zeeman splitting induced in the semiconducting region. On the other hand, the configuration of Fig. 1(b) seems to be fully compatible with both N metals and semiconductors, thanks to the spatial localization of $h_{\text{exc}}$ provided by the FS bilayer \cite{Bergeret}. In this context,  promising F candidates are represented by soft ferromagnetic alloys such as Pd$_{1-\alpha}$Ni$_\alpha$ \cite{kontos} or Cu$_{1-\alpha}$Ni$_\alpha$ \cite{ryazanov}, which provide tunable $h_{\text{exc}}$ thanks to a proper choice of $\alpha$.
As far as the structure realization is concerned, such ballistic junctions could be fabricated either by making a mechanical contact between  a sharp metallic tip and the surface of a thin film (i.e., the point contact technique) \cite{blonder,yanson}, or by defining nanoscale holes (3-20 nm diameter) in silicon-nitride insulating membranes, as reported in Refs. \cite{ralls,upadhyay,chalsani}.

We stress, finally, that superconductors dominated by Zeeman energy possess an additional characteristic. Namely, the presence of the gap allows efficient quasiparticle \emph{cooling} \cite{RMP} in the N region for properly-tuned low-transmissive NS interfaces (i.e., $Z\gg1$) \cite{GiazottoPRB2007}.  
This may yield a significant enhancement of $P$ upon current injection at finite $T$ (see Fig. 4).
By contrast, injection from a F may only lead to \emph{heating} of the electron gas \cite{RMP}. 

In summary, we have demonstrated that highly spin-polarized currents are obtained in NS junctions when the superconductor is dominated by Zeeman interaction.
The sign and magnitude of polarization can be tuned by varying bias voltage and Zeeman energy.
While for a tunnel contact 100 \% polarization is attainable, for a transparent interface $P$ is dominated by the minority spin species, in contrast to what happens for a ferromagnetic injector.
All this makes superconductors ideal and easy-accessible spin sources for low-temperature spintronics.

Partial financial support from the NanoSciERA "NanoFridge" and RTNNANO projects of the EU is acknowledged.


\begin{thebibliography}{99}
\bibitem{proximity}
	For an opaque contact, the superconductor is marginally affected by the proximity of the F layer.
\bibitem{parameters}
	For our calculations we set $\Delta_0=1$ meV,	$\mu=5$ eV, $g=2$, and $\mathcal{A}=2.5\times 10^{-3}\,\mu$m$^2$.
\end{thebibliography}
\end{document}